\RequirePackage{fix-cm}
\documentclass[smallextended]{svjour3}       
\smartqed  
\usepackage{graphicx}
\begin{document}

\title{CONCERTO at APEX: installation and technical commissioning} 



\author{
A. Monfardini \and
A.~Beelen \and
A.~Benoit \and
J.~Bounmy \and
M.~Calvo \and
A.~Catalano \and
J.~Goupy \and
G.~Lagache \and
P.~Ade \and
E.~Barria \and
M.~B\'ethermin \and
O.~Bourrion \and  
G.~Bres \and
C.~De Breuck \and
F.-X.~D\'esert \and
G.~Duvauchelle \and
A.~Fasano \and  
T.~Fenouillet \and
J.~Garcia  \and 
G.~Garde \and 
C.~Hoarau \and 
W.~Hu \and
J.-C.~Lambert \and
F.~Levy-Bertrand \and
A.~Lundgren \and
J.~Macias-Perez \and
J.~Marpaud \and
G.~Pisano \and
N.~Ponthieu \and
L.~Prieur \and
S.~Roni \and
S.~Roudier \and
D.~Tourres \and
C.~Tucker \and
M.~Cantzler \and
P.~Caro \and
M.~Diaz \and
C.~Dur\'an \and
F.~Montenegro \and
M.~Navarro \and
R.~Olguin \and
F.~Palma \and
R.~Parra \and
J.~Santana 
}


\institute{A.Monfardini, M.~Calvo, J.~Goupy, J.~Bounmy, A.~Benoit, E.~Barria, O.~Bourrion, G.~Bres, G.~Garde, C.~Hoarau, F.~Levy, J.~Macias-Perez, J.~Marpaud, S.~Roni, S.~Roudier, D.~Tourres \at
Institut Néel and LPSC, CNRS and UGA, 25-53 rue des Martyrs, Grenoble, France
\and
A.~Beelen, G.~Lagache, M.~Bethermin, G.~Duvauchelle, A.~Fasano, W.~Hu,  T.~Fenouillet, J.~Garcia, J.-C.~Lambert, L.~Prieur 
\at
Aix Marseille Universit\'e, CNRS, LAM, F-13388 Marseille, France
\and
P.~Ade, G.~Pisano, C.~Tucker 
\at
AIG, University of Cardiff, The Parade, CF24 3AA, United Kindgom
\and
C.~Duran, M.~Cantzler, P.~Caro, R.~Olguin, R.~Parra, M.~Diaz, F.~Montenegro, F.~Palma, J.~Santana, M.~Navarro
\at
European Southern Observatory, Alonso de Cordova 3107, Vitacura, Santiago, Chile
\and
C.~De Breuck, A.~Lundgren
\at
European Southern Observatory, Karl Schwarzschild Straße 2, 85748 Garching, Germany
}

\date{Received: date / Accepted: date}

\maketitle

\begin{abstract}
We describe the deployment and first tests on Sky of CONCERTO, a large field-of-view (18.6\,arc-min) spectral-imaging instrument. The instrument operates in the range 130-310\,GHz from the APEX 12-meters telescope located at 5100\,m a.s.l. on the Chajnantor plateau. Spectra with R=$\nu / \Delta \nu \leq 300$ are obtained using a fast (2.5\,Hz mechanical frequency) Fourier Transform Spectrometer (FTS), coupled to a continuous dilution cryostat with a base temperature of 60\,mK. Two 2152-pixels arrays of Lumped Element Kinetic Inductance Detectors (LEKID) are installed in the cryostat that also contains the cold optics and the front-end electronics. CONCERTO, installed in April 2021, generates more than 20k spectra per second during observations. We describe the final development phases, the installation and the first results obtained on Sky.
\end{abstract}

\section{Introduction}
\label{intro}
CONCERTO is a millimeter-wave low spectral resolution (R=$\nu / \Delta \nu \leq 300$) imaging-spectrometer with an instantaneous field-of-view of 18.6\,arc-min. Thanks to the combination of high mapping speed, relatively high angular resolution ($\approx$30\,arc-sec) and spectral capabilities, CONCERTO is opening new observational windows of the millimetre sky. It employs a room-temperature Martin-Puplett Interferometer (MpI) \cite{mpi1970} coupled to a large field millimeter-wave camera, and has been designed to interface with the Atacama Pathfinder EXperiment (APEX) 12-meters telescope located on the Chajnantor plateau \cite{gusten2006}. The detectors technology is derived from the one developed for NIKA \cite{monfardini2011}, NIKA2 \cite{adam2018} and KISS \cite{fasano2020}. We have described the CONCERTO design and science forecast in detail in a previous publication \cite{concerto2020}. In the present paper we will focus on the description of the final configuration, the installation at the telescope and the technical commissioning.

\section{Instrument description}
\label{sec:instrument}

CONCERTO has two main components: the so-called "chassis" and the "optics box". The \emph{chassis} includes the camera (dilution cryostat), the MpI interferometer, the readout and control electronics. The \emph{optics box} includes seven room-temperature mirrors (M5 to M11), two polarisers and a cold reference (with three additional mirrors) for the MpI. The camera employs refractive optics, with three lenses (300\,K, 4\,K and 0.1\,K) and a third polariser (0.1\,K). The M3 and M4 mirrors, interfacing CONCERTO to the telescope, are mounted respectively on top of the chassis and on the ceiling of the APEX Cassegrain cabin (C-cabin). The M3 mirror is remotely foldable.

The \emph{dilution cryostat} has been entirely designed and fabricated in house. In order to account for the telescope elevation changes during observations, the dilution insert is specially conceived to work at inclinations up to 75 degrees, corresponding to a telescope elevation comprised in the range 15-90 degrees. The cryostat is completely cryogen-free, including the cold trap for $^3$He injection. The \emph{cold reference} is a second cryostat with a base temperature of roughly 30\,K. The main cryostat and the cold reference are both based on pulse-tubes. Due to installation constraints, the two pulse-tubes must share a single helium compressor and one pair of pipes. The cold-heads spinning motors are properly synchronised by an ad-hoc electronics box developed in house. This configuration doesn't significantly affect the performance of the dilution cryostat, but limits the base temperature of the cold reference. In a future upgrade, it is planned to replace the helium compressor and reduce the temperature of the cold reference to 10\,K roughly. The duty cycle of the CONCERTO cryogenics is 100\,\%, i.e. continuous operations. During the technical commissioning, the cryostats have been cold and stable for 1.5\,months without interruptions.

The two \emph{focal planes} (HF and LF, separated by a 45 degrees polariser) are single polarisation LEKID arrays containing 2152 pixels each. The focal plane is a circle of 76\,mm in diameter, while the pixel size is 1.4$\times$1.4\,mm$^2$. The LEKID are microstrip-coupled and based on high-quality thin (20\,nm) Aluminium films evaporated on high-resistivity mono-cristalline Silicon substrates. The thickness of the substrates are respectively 105\,$\mu$m and 125\,$\mu$m for the HF and LF focal planes. This allows to cover the desired optical bands, i.e. 195-310\,GHz for the HF and 130-270\,GHz for LF. On the back of the wafer, a thicker (200\,nm) layer acts as superconducting ground plane, while an additional Au film (100\,nm) is added to ensure a good thermalisation onto the gold-plated Copper holder. Each of the twelve readout lines (six per array) is connected to a custom readout board, with a multiplexing factor of 400 over a bandwidth of 1\,GHz. The band-defining optical filters are mounted in front of the arrays, and consist in a pair of low-pass and high-pass metallic multi-mesh. 

\begin{figure*}[ht]
  \includegraphics[width=1.0\textwidth]{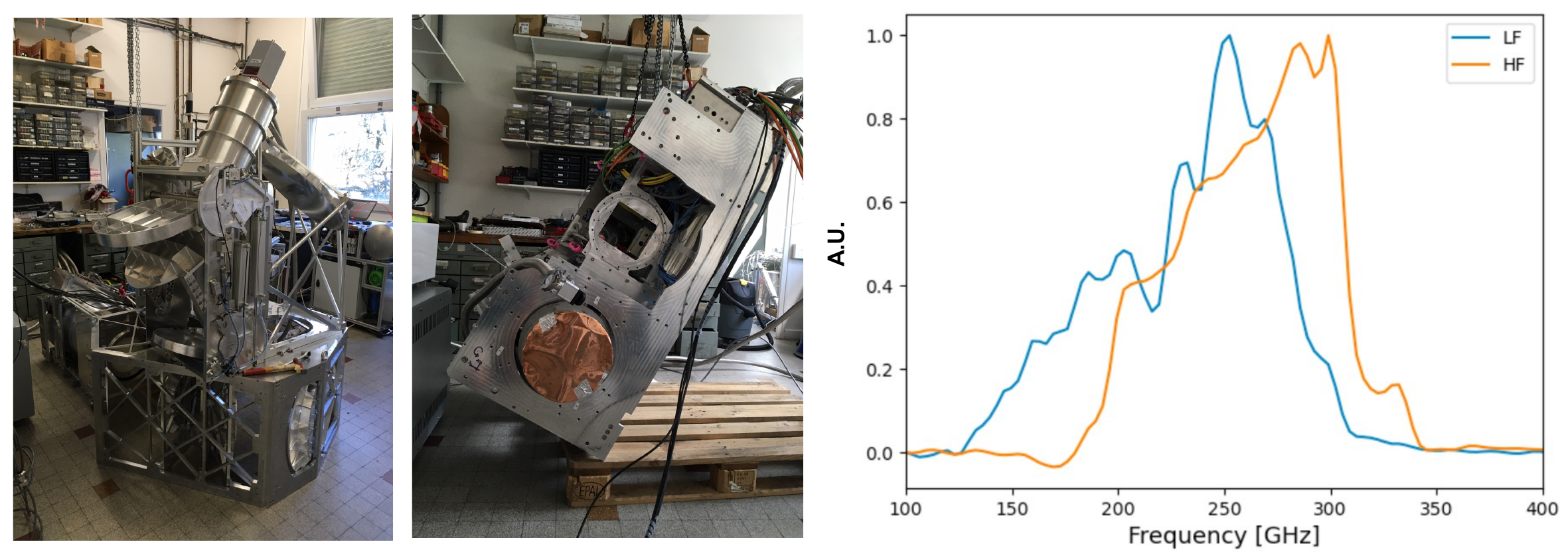}
\caption{\emph{Left:} CONCERTO integrated in laboratory. \emph{Center:} chassis inclination test. \emph{Right:} raw band-passes measured in laboratory.}
\label{fig:labo}       
\end{figure*}

In this section we have been focusing mainly on the sub-systems that have been finalised after our design paper \cite{concerto2020}. Please refer to that publication for a more detailed description of the overall architecture of the instrument.

%

\section{Installation at the telescope}
\label{sec:installation}
The instrument has been fully integrated and tested, excluding the M3 mirror, in laboratory. In particular, we have used our Sky Simulator \cite{monfardini2011} to demonstrate spectral-imaging capabilities. In February, 2021, the instrument was dismounted and shipped to Chile. The installation, submitted to severe COVID-related restrictions, started on the 6$^{th}$ of April, 2021. The elements in the C-cabin were installed in the following order: 1) M4; 2) optics box base; 3) chassis; 4) cold reference optics and cryostat; 5) M3. 
The pulse-tube compressor, the dilution Gas Handling System (GHS) and the He isotopic mixture reserve are installed in the lower part of the telescope tower, in the so-called "compressors room". The internal network switch, and the data acquisition (DAQ) computers, linked to the readout electronics by four dedicated 1Gb Ethernet cables, are located in an intermediate container. The real-time analysis machine and the cold storage are installed in the servers rooms, few tens of meters away from the telescope, linked with two redundant 10Gb Ethernet cables. CONCERTO is controlled remotely; no manual intervention is needed during normal operations, including cooling down and warming up. 

\begin{figure*}[ht]
  \includegraphics[width=1\textwidth]{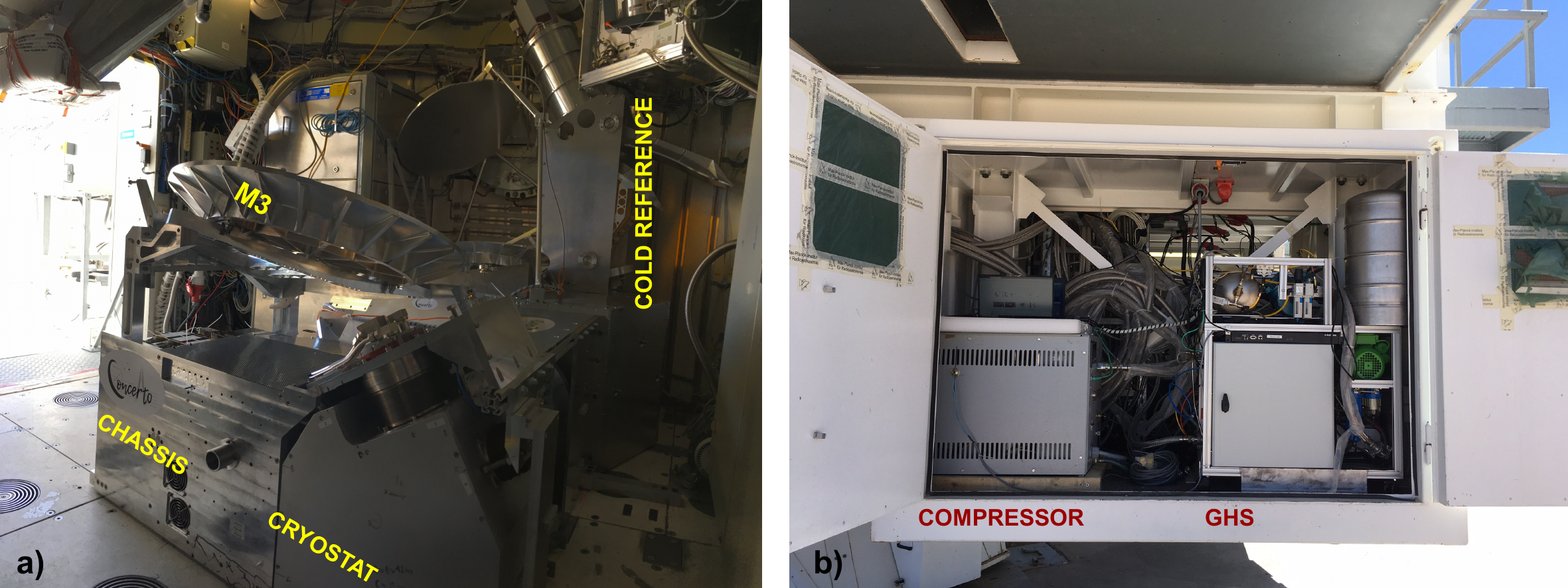}
\caption{Picture of CONCERTO, installed at APEX, April 2021. \emph{Left:} the instrument in the C-cabin, including the chassis, the optics, the cold reference and the M3 mirror. \emph{Right:} after 23 meters of cables and pipes, in the compressors room, the pulse-tubes compressor, the Gas Handling System (GHS) and the $^3$He-$^4$He 45 liters reserve.}
\label{fig:installation}       
\end{figure*}

\section{Technical commissioning}
\label{sec:commissioning}

\begin{figure*}[ht]
  \includegraphics[width=1.0\textwidth]{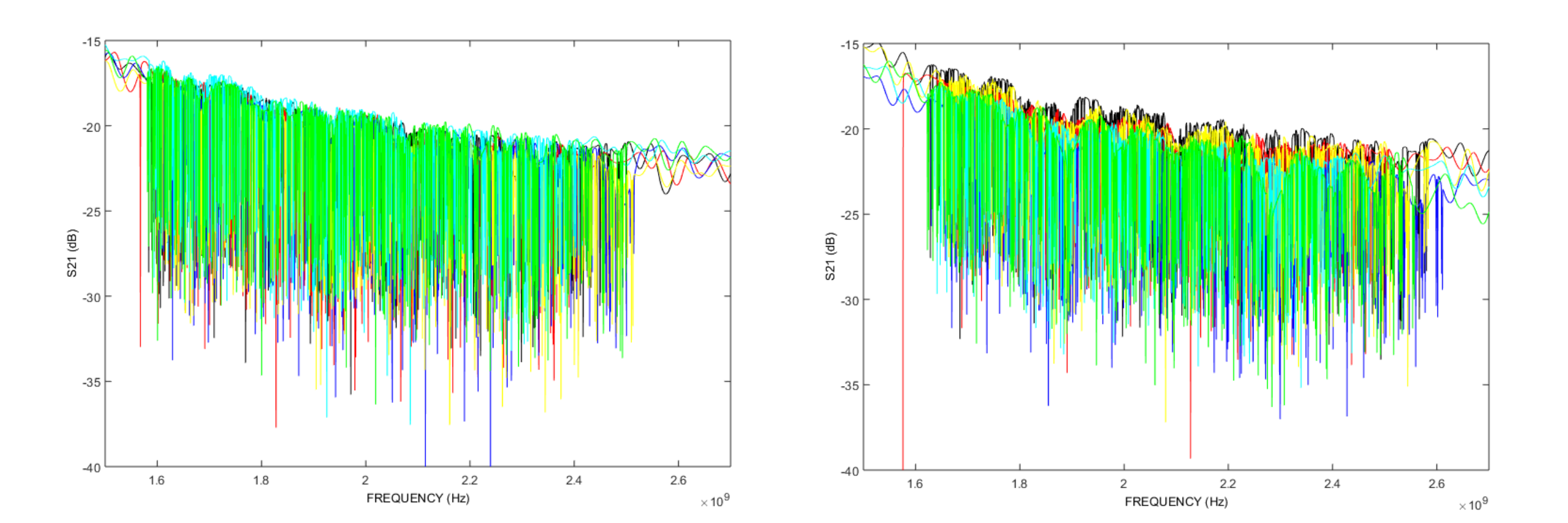}
\caption{Frequency sweeps of the two arrays. More than 90\% of the 4,304 designed pixels exhibit a resonance. \emph{Left:} LF array. \emph{Right:} HF array.}
\label{fig:VNA}       
\end{figure*}

The first cool-down of CONCERTO at the telescope started on the 10th of April, 2021. It was performed while the telescope was used for heterodyne observations, i.e. re-pointing frequently. The base temperature was reached in roughly 48 hours, as during laboratory tests. The integrity of the arrays and cold electronics was first verified by acquiring frequency sweeps on the twelve readout lines, as shown in figure \ref{fig:VNA}. All the lines were connected, and showing nominal resonances. On top of that, we observed an improvement of the LEKID internal quality factors ($Q^i$) with respect to the laboratory measurements. This is due to the superior quality of the Sky at Chajnantor with respect to the Sky Simulator. We have in particular $Q^i_{Sky-Simulator} \approx 10k$ versus $Q^i_{Sky-APEX} \approx 17k$. The internal quality factors measured on Sky match perfectly with the (designed) coupling quality factors $Q^c \approx 15k$. This is, according to our past experiences, a necessary condition to achieve optimal performance. 

The optics response on Sky is of the order of 1\,kHz resonance shift per K of temperature background. The noise was initially measured by targeting the Sky without moving the telescope from the zenith position. The continuum component, in the useful frequency band, is of the order of 2.5\,Hz/$\sqrt{Hz}$. The noise in our case is expressed in terms of the localisation precision of each resonance in frequency (Hz) \cite{calvo2013}. That translates in a Noise Equivalent Temperature (NET) of about 2.5\,mK/$\sqrt{Hz}$/pixel/array. By combining the two arrays, and taking into account that each beam is sampled by more than two pixels, we should in the end achieve NET $\leq$ 1\,mK$\cdot \sqrt{s}$ per beam. These values are in agreement with our initial sensitivity forecasts \cite{concerto2020}.

\begin{figure*}[ht]
  \includegraphics[width=1.0\textwidth]{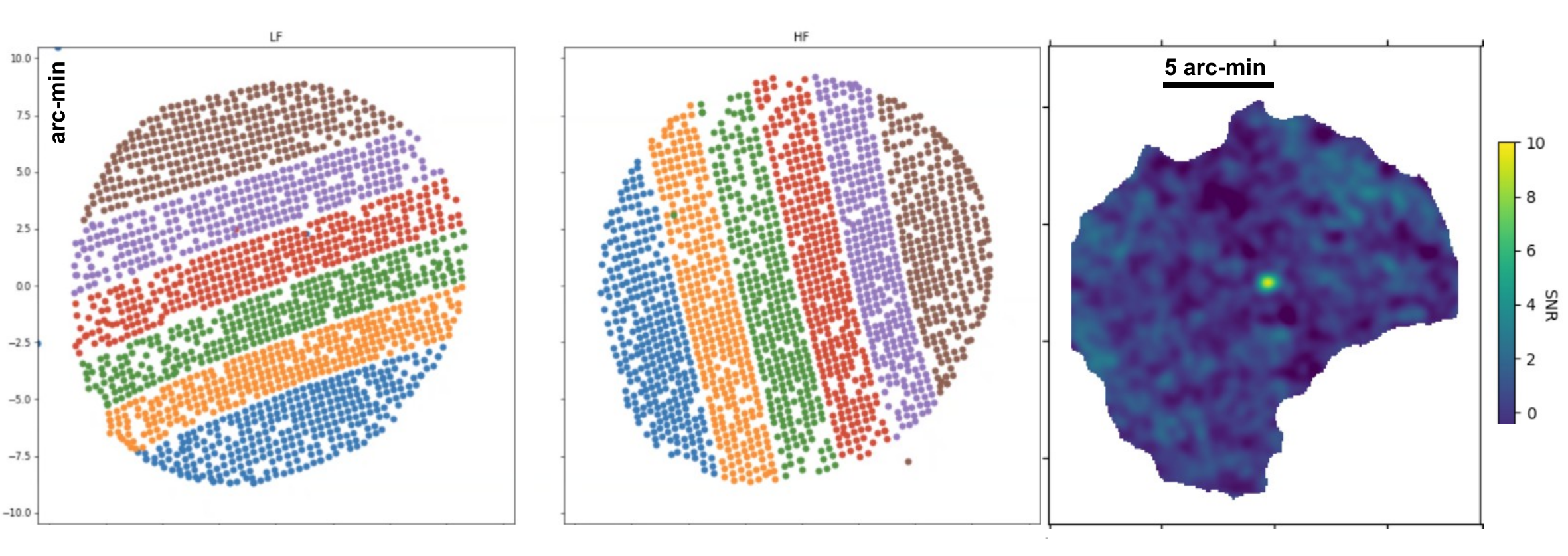}
\caption{\emph{Left and Center:} LF and HF focal plane geometries. Around 90\% of the 4,304 designed pixels exhibit a beam. 
\emph{Right:} LF uncorrected quick-look pointing map ($t_{telescope} \approx$\,35\,sec) of the J0609-157 AGN (ALMA flux $\approx$\,1.3\,Jy at 233\,GHz). Only the pixels exhibiting an ellipticity beam lower than 0.7 are projected.
}
\label{fig:geometry}       
\end{figure*}

The next step in the technical commissioning was to set up CONCERTO in the final observing condition, that includes software and data format compatibility with the APEX observatory. During these sessions, second half of April 2021, we have performed a number of observations, mostly technical routines like focus, pointing, sky dips, interferometer tests. We report in figure \ref{fig:geometry} an example among the very first results. We focus for now, due to space constraints in these proceedings, on the photometric mode. In particular, we show the projected geometries of both arrays, and a pointing map. This last demonstrates that CONCERTO is able to map sub-Jansky features, over hundreds of arc-min$^{2}$, in less than one minute overall telescope time.  

Taking into account the convolution of the pixel footprint, i.e. a square of 22$\times$22\,arc-sec$^2$ when projected on Sky, and the resolution of the telescope, we demonstrate a good agreement with the observed beams FWHM, i.e. in the range of 30(HF)-35(LF)\,arc-sec. Besides the distortions in the overall shape of the focal plane shown in figure \ref{fig:geometry}, we observe a degradation of the beams quality, in particular in terms of ellipticity, on the same side of both arrays. Around 25\% of the pixels are affected by this anomaly, as shown in the right panel of figure \ref{fig:geometry}. These effects, most likely of optical origin, are under investigation and characterisation. Similar phenomena have been observed by large field-of-view instruments previously installed at APEX \cite{siringo2009,schwan2011}.

\section{Conclusions}
\label{sec:conclusions}

The wide field-of-view low-resolution spectrometer CONCERTO is, since April 2021, installed and operating at APEX. We have described the final configuration, and presented the results of the installation and technical commissioning. The instrument is fully functional, with 90$\%$ of the 4,304 pixels exhibiting beams. The preliminary sensitivity estimations are well in agreement with expectations. As a result, we are planning a science verification campaign during the Summer 2021. Among the effects being investigated further: 1) a monochromatic optical noise at a frequency of tens of Hz and coming from the C-cabin; b) the stability of the interferometer; c) some unexplained optics aberrations. During the technical commissioning, we have also observed a number of fainter and extended sources. Despite preliminary, these observations are clearly showing the big potential of CONCERTO in terms of mapping speed. These maps, and further results including spectro-photometric observations, will be presented in future publications.

\begin{acknowledgements}
We thank the electronics, mechanics, cryogenics and administrative groups at Institut N\'eel, LPSC and LAM. We thank Florence Picchiottino, Karina Celedon and Lilia Todorov for taking care of the shippings. For the outstanding technological and human support we thank the whole APEX staff. Among them, the telescope operators Juan Pablo Perez, Edouard Gonzalez, Mauricio Martinez. Thank you all, it was a unique experience ! We acknowledge financial support from the European Research Council (ERC grant 788212), the Aix-Marseille University-A*Midex and the LabEx FOCUS.

\end{acknowledgements}

%
%



\end{document}